# Improving Linewidth and Extinction Ratio Performances of Lithium Niobate Ring Modulator Using Ring-pair Structure

*Songyan Hou*

**Electro-optic modulators lie at the heart of complex integration and high density electro-optic systems. One of the representative electro-optic modulators is thin film lithium niobate based microring modulator which has demonstrated advantages of compact footprint, low optical loss and high modulation efficiency. However, the linewidth and extinction ratio of ring modulators are fundamentally limited by the ring losses and coupling, respectively. To this end, we propose a novel type of electro-optic modulators with ring-pair structure on thin film lithium niobate platform, which brings substantially improvement of linewidth and extinction ratio. The ring-pair modulator exhibits a larger linewidth up to 22 GHz, 1.74-time larger than that of single ring resonator with same design parameters. Moreover, the experimental results also reveal that the added-up extinction ratio of ring-pair resonator goes beyond 30 dB, much larger than that in an individual ring resonator. These advantages of ring-pair modulator pave a new way for the application of compact ring-based modulators with large working wavelength window and high extinction ratio, to be exploited in quantum optics, programmable nanophotonics and optical sensors, etc.**

## 1. Introduction

Electro-optic modulators (EOMs) convert high speed electronic signals into optical signals. Such devices are of paramount significance to practical applications in telecommunication networks[1-3], microwave signal processing[4, 5] and emerging applications such as quantum information[6] and optical sensors[7, 8], etc. The migration of EOMs to integrated devices is motivated by the demand for better performance modulators with large bandwidth, low drive voltage, high extinction ratio, compact footprint and the compatibility with large scale integration. These requirements have resulted in the rapid development of EOMs based on many photonic platforms, including silicon[9], indium phosphide[10], discrete lithium niobate (LN)[11, 12], and plasmonics[13]. While the EOMs have been greatly improved in the above platforms, the realization of a modulator that concurrently satisfies all necessary performance features is still challenging due to the limited electro-optic performance of the platform materials. For example, silicon is known as a significant photonics platform owing to its excellent scalability and mature CMOS fabrication infrastructure. However, the implement of electro-optic effect in silicon heavily relies on the doped free carrier, and the lack of intrinsic electro-optic effect of silicon prevents its further applications for high-speed systems with required bandwidth and low power consumption.

Recently, the thin film LN on insulator platform has emerged as a practical solution for high performance electro-optic photonic devices[14-18]. Importantly, the thin film LN platform combines the excellent electro-optic properties of bulk LN with advantages of photonic film such as high index contrast and tightly confined modes[19-29]. Although high performance EOMs systems with low drive voltage and ultrahigh bandwidth have been experimentally demonstrated using Mach–Zehnder interferometer (MZI) on thin film LN[24, 27, 30], the centimeter length of the MZI devices presents a big challenge for the applications in large-scale integrated optical interconnection systems.

To further reduce the footprint of a modulator, resonant ring modulators have been widely studied. Ring modulators are key components in short-range optical interconnects due to their low operating voltage, compact size, and compatibility with CMOS circuit drivers. Micron-sized electro-optic modulators based on microring resonators presents great advantages in interconnection systems with dense wavelength division multiplexing (DWDM)[31, 32]. The driving voltages of the microring modulators are generally low since the quality factor (Q-factor) of microring resonator can be extremely high with narrow linewidth[33]. Even so, the large-scale applications of microring modulators are hindered by the following two aspects. First, the extinction ratio of microring modulators suffers from its sensitivity to the fabrication variations and environment perturbation. This is because that extinction ratio stringently depends on the coupling condition between bus waveguide and ring. In particular, the highest extinction ratio occurs only under critical coupling, which can be readily destroyed by any slight external perturbation. Such a sensitivity makes it difficult to realize the microring modulators with a high signal-noise ratio. Secondly, the low-loss microring modulator is generally accompanied with a narrow working linewidth. This effect results from the fundamental trade-off between the Q-factor and linewidth.

In this paper, we propose a high-performance ring-based EOM using ring-pair resonators on the thin film LN platform, i.e., the ring-pair thin film LN modulator. We highlight that our design not only sustains the superior electro-optic properties of LN, but also enhances the working wavelength window and extinction ratio simultaneously. To be specific, our experimental results successfully demonstrate that the fabricated ring-pair thin film LN modulator exhibits a large working wavelength range (i.e., up to 0.1777 nm) with high Q and high extinction ratio (i.e., > 30 dB) without the necessity of critical coupling. Such a linewidth is 1.74-time wider than that of single-ring LN modulator without sacrificing the Q-factor. Moreover, the extinction ratio of the ring-pair thin film LN modulator is also enhanced by around 2 times as compared to the conventional modulator based on individual ring resonator. All these features make our device a high-performance EOM that is compact, broadband and highly efficient.

*2. Principle and Design*

**Figure 1a** shows the schematic diagram of the designed ring-pair thin film LN modulator. The ring-pair thin film LN modulator comprises of a bus waveguide, two identical microring resonators and only one pair of Signal-Ground electrodes. The LN waveguides are designed with their width to be 1.5 um and sit on top of a 200 nm LN slab. As such, the waveguides can support fundamental transverse electric modes at the wavelength around 1550 nm. The two identical microring resonators are modulated by a pair of Signal-Ground electrodes but experience inverse electric fields (**Figure 1b** and **1c**). In addition, the whole devices are cladded with 1.5 um silicon dioxide ($SiO_2$) layer to avoid the unwanted optical absorption from the top metallic strips (see more details in the section of **Methods**).

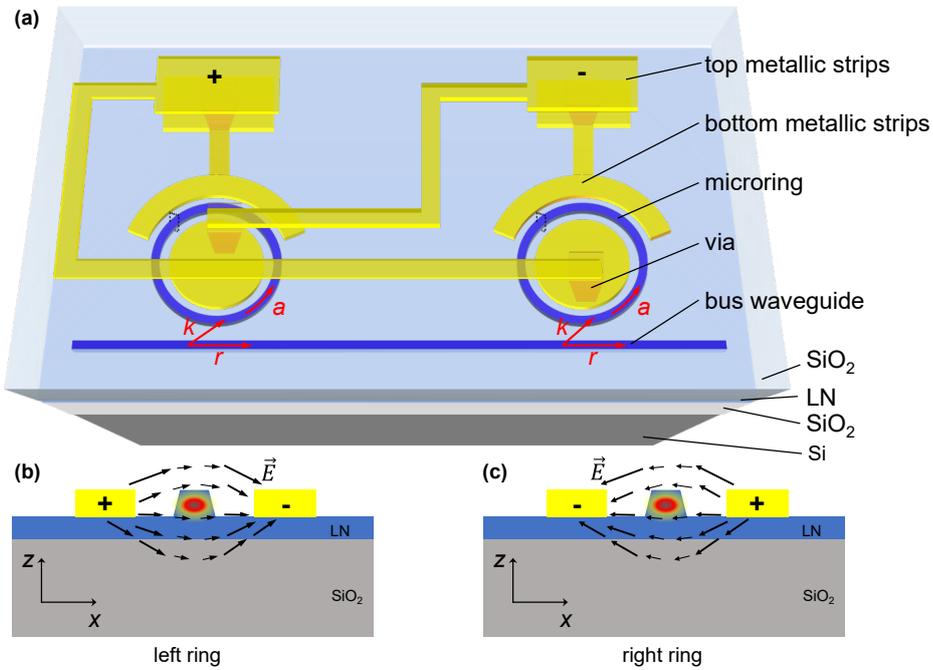

*Figure 1*. a) Schematic diagram of the thin film LN ring-pair modulator and electrical contacts. *r* represents the self-coupling coefficient at coupler; *k* is the cross-coupling coefficient and *a* represents the single pass transmission amplitude. b,c) Cross-section view of TE mode in LN waveguides in left (*b*) and right (*c*) microring at marked position in *a*. The two arrows show the direction of applied electric field.

To evaluate the linewidth of the thin film LN modulator, we calculate the transmission spectrum of a single-ring resonator. Under the assumption of negligible backward reflection into the bus waveguide, the transmitted optical field normalized to the incidence can be obtained as:

$$\frac{E_{pass}}{E_{input}} = e^{i(\pi+\phi)} \frac{a - re^{i\phi}}{1 - are^{i\phi}} \tag{1}$$

Here, $\phi = \beta L = \frac{2\pi n L}{\lambda}$ is the phase shift after single pass in microring, in which $\beta$ is the propagation constant, $n$ is the refractive index and $L$ is the ring round length. $a$ represents the single pass transmission amplitude, including coupling loss and propagation loss in waveguide. It is determined by the optical power attenuation coefficient $\alpha$ as: $a^2 = e^{-\alpha L}$. $r$ represents the self-coupling coefficient at coupler.

The full width at half maximum (FWHM) at resonance can be derived as:

$$\Delta w_{single} = \frac{(1-ra)\lambda_{res}^2}{\pi n L \sqrt{ra}} \tag{2}$$

Here, the $\Delta w_{single}$ of single ring is positively correlated with the optical loss $\alpha$, which thereby leads to a fundamental trade-off and link between working linewidth and optical loss in single ring modulator.

For a ring-pair resonator shown in **Figure 1a**, given that the two rings are identical, the transmission of a ring-pair resonator can be derived as:

$$T_{ring-pair} = (T_{single})^2 = \left(\frac{a^2 - 2ra\cos\phi + r^2}{1 - 2ra\cos\phi + (ra)^2}\right)^2 \tag{3}$$

Clearly, the FWHM of ring-pair resonator can be obtained from **Equation 3** as:

$$\Delta w_{ring-pair} = \frac{(1-ra)\lambda_{res}^2}{\pi n L \sqrt{(\sqrt{2}-1)ra}} \tag{4}$$

Compared to single ring resonator, the **Equation 2** and **Equation 4** show that the ring-pair resonator demonstrates a larger linewidth without increasing the optical losses: $\Delta w_{ring-pair} = \sqrt{(\sqrt{2}+1)} \Delta w_{single} \approx 1.55 \Delta w_{single}$. This increased linewidth of ring-pair resonator is further illustrated in the simulated transmission spectra in **Figure 2a** with microring parameters indicated in the legend.

On the other hand, the transmission equations also help us to investigate the extinction ratio of resonators, which is another key parameter of an optical modulator. The extinction ratio of resonators is defined as the ratio of minimum transmission $T_{min}$ and maximum transmission $T_{max}$. Providing that the $T_{max} \approx 1$, the extinction ratios of single-ring and ring-pair resonators can be derived as:

$$ER_{single} = 10 \times \log_{10}(T_{min\text{-}single}) \tag{5}$$

$$\begin{aligned} ER_{ring-pair} &= -10 \times \log_{10}\left(\frac{T_{max\text{-}(ring-pair)}}{T_{min\text{-}(ring-pair)}}\right) \\ &= 10 \times \log_{10}(T_{min\text{-}single})^2 \\ &= 2ER_{single} \end{aligned} \tag{6}$$

**Equation 5** and **Equation 6** show that the extinction ratio of a ring-pair resonator is doubled under same parameters of the single-ring resonator, as shown in **Figure 2a**.

To further analyze the modulation performance of single-ring and ring-pair modulators, the optical response to applied voltage is simulated in **Figure 2b**, **2c** and **2d**. The transmission spectra of the ring-pair modulator show a peak in the center of a broader dip when voltage applied (**Figure 2c**), which is different with the single-ring modulator transmission spectra (**Figure 2b**). The transmission of center peak is dependent on voltage introduced inverse resonance detuning between the two rings, which is used for amplitude modulation (**Figure 2c** and **d**).

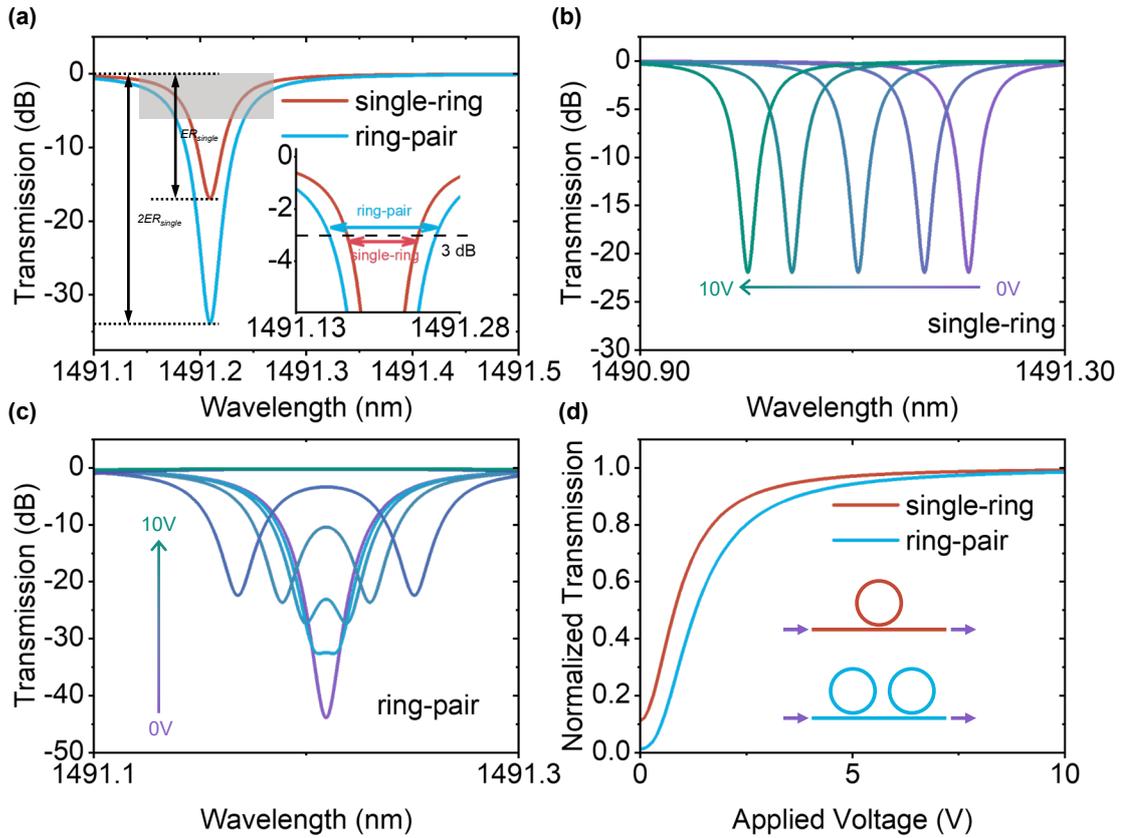

*Figure 2*. Simulated transmission spectra of single-ring and ring-pair modulators with ring radius: 10 um, self-coupling coefficient r=0.998, optical loss a=0.996. a) The transmission spectra of resonators at 0 V. The inset shows enlarged plot at grey area. b) The resonance shifts of single-ring resonator with applied voltages. c) Transmission spectra of ring-pair resonator with various voltages. d) The transmission intensity versus applied voltages.

## 3. Experimental results

We fabricated the ring-pair modulators with radius of 40 um on x-cut thin film LN platform (**Figure 3a**), as described in **Methods** section. To ensure the highest electro-optic effect in LN, a fundamental transverse-electric (TE) mode is used as pump mode. The extinction ratios of two rings are 19 dB and 13 dB ( $ER_{ring2}$ and $ER_{ring1}$, **Figure 3b**), respectively. Apparently, the two rings are under coupled (usually the ER of critical coupling is larger than 20 dB). When the resonances of two rings are well aligned (**Figure 3c**), the ring-pair resonator shows 32 dB extinction ratio with $ER_{ring\text{-}pair} = ER_{ring1} + ER_{ring2}$. In addition to higher extinction ratio, the ring-pair structure demonstrates a larger 3-dB linewidth (22 GHz) with $\frac{BD_{double}}{BD_{single1}} \approx 1.28$ and $\frac{BD_{double}}{BD_{single2}} \approx 1.78$. These experimental results match well with the simulations shown in above. The intrinsic Q-factors of two rings are 20k and 25k, respectively, and the optical power coupling coefficient between rings and bus waveguides are calculated as 7.1% and 4.5%, respectively. These performance variances of the two rings come from the fabrication discrepancy.

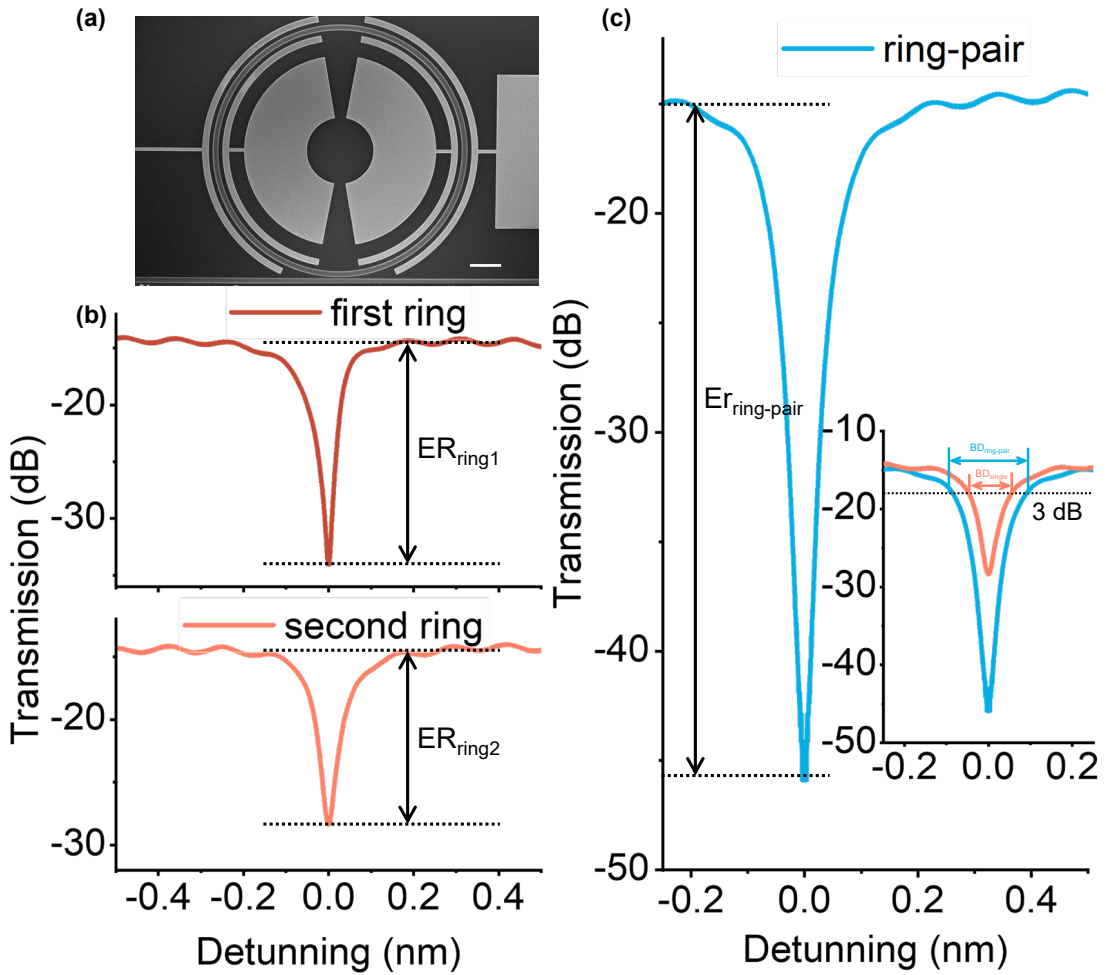

***Figure 3***. *Optical characterization of single-ring and ring-pair resonator without detuning. a) Scanning electron microscope image of the fabricated ring resonator with electrodes. Scale*

*bar: 10 um. b) Transmission spectra of first ring and second ring with indicated extinction ratios. c) Transmission spectrum of ring-pair resonator. Inset shows the linewidth comparison between ring-pair and second ring resonators.*

To further compare the modulation performance between single-ring and ring-pair modulators, electric-optic response of single-ring and ring-pair modulators are measured (see details in **Methods**). **Figure 4a** shows the transmission responses of single ring modulator when applied with various voltages. The effective refractive index of ring path changes with voltages, which leads to the linear shift of resonance wavelength (**Figure 4b**). The tunning efficiency of single-ring modulator is 4.8 pm/V without obvious fluctuation in extinction ratio and linewidth (**Figure 4a** and **b**). **Figure 4c** shows the modulation performance of ring-pair modulator. At 0 V, the resonant wavelengths of the two rings are 1561.33 nm and 1561.39 nm respectively. The peak in the center of transmission spectra in ring-pair modulator results from the multiplication of the two single rings. The optical modulation occurs when applied with voltages, which induce opposite electric fields within two rings and thereby lead to the resonance frequencies of the two rings driven to opposite directions (**Figure 4c**). The ring-pair modulator works with large linewidth and high extinction regardless of the initial resonance offset between the two rings, largely relaxing the fabrication precision. When the resonant frequencies in two rings are aligned, the center peak disappears, and extinction ratio adds up to around 32 dB (**Figure 4c**). Owing to the transmission multiplication effect, the transmission of ring-pair modulator increases slower around minimum and maximum transmission, which leads to the modulation efficiency of ring-pair modulator ($\Delta V = 55V$, **Figure 4c**) is slightly larger than that of single-ring modulator ($\Delta V = 50V$, **Figure 4a**). Note that the modulation efficiency can be improved by using racetrack resonators or increasing the ring radius and the micro-structured electrode design can be used to boost the modulation bandwidth. To demonstrate the advantages of our design, we compare our device performance with that of previously reported LN resonators (see Table 1): First, it is obvious that there is a trade-off between linewidth and Q-factor of the ring: the linewidth dramatically shrinks when the Q-factor increases. Our monolithic LN ring-pair modulator demonstrates a large linewidth of 0.177 nm with a high Q-factor around 25k. Second, the extinction ratio of ring resonators is limited by the coupling with a typical value around 15 dB. The highest extinction ratio happens only at critical coupling, which is a big challenge for practical applications. Our ring-pair device demonstrates over 30 dB extinction ratio with two rings under coupled. Moreover, our device has a high tunning efficiency of 4.8 pm/V with a small radius footprint of 40 um.

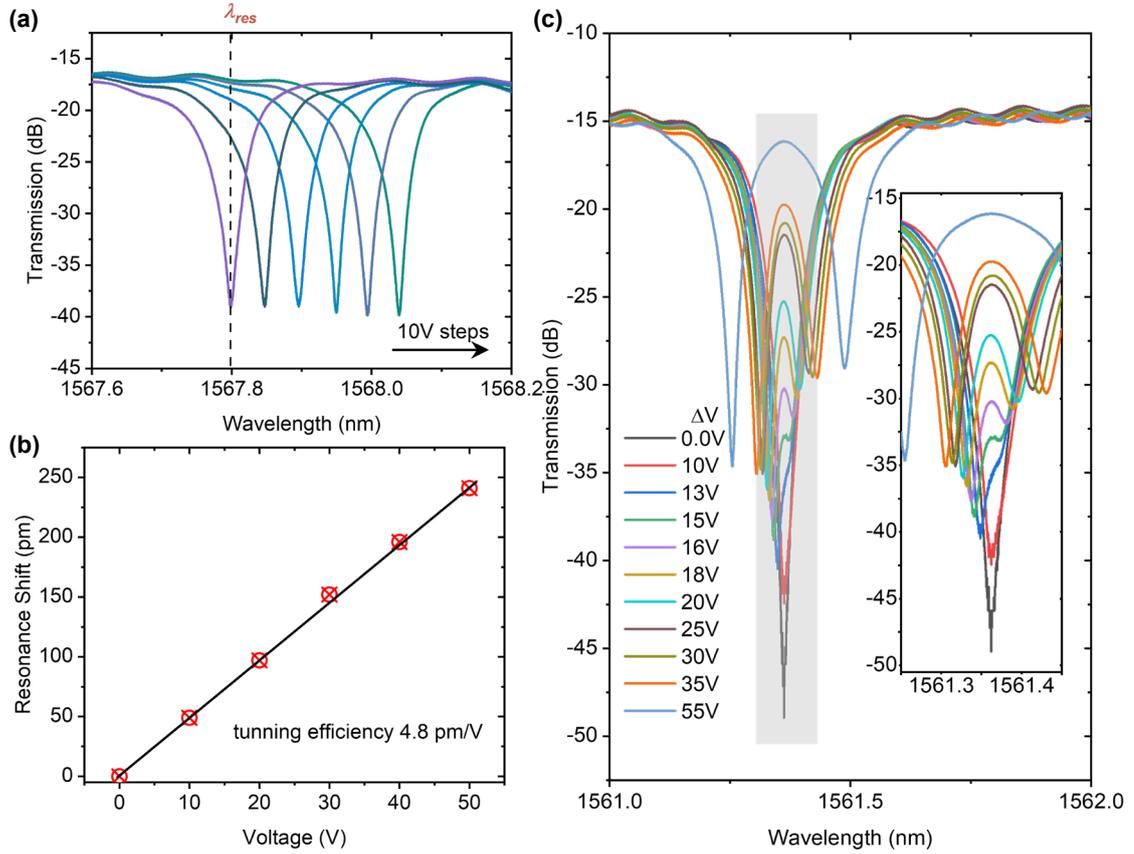

*Figure 4*. Modulation performance of single-ring and ring-pair modulators. a) Measured transmission spectra of single-ring modulator with various voltages. b) Linear resonance shift of single-ring modulator as a function of applied voltage. c) Measured transmission spectra of ring-pair modulator with various voltages.

| Resonators | Linewidth (nm) | Q (× 10³) | Radius (um) | Tunning efficiency (pm/V) | ER (dB) |
|---|---|---|---|---|---|
| Monolithic Ring[34] | 0.25 | 7.5 | 70 | 3 | 3 |
| Monolithic Ring[35] | 0.005 | 800 | 70 | -- | 15 |
| Hybrid Ring[36] | 0.1 | 14 | 15 | 3.3 | 10 |
| Hybrid Ring[37] | 0.015 | 120 | 200 | 3.2 | 15 |
| This work | 0.1777 | 25 | 40 | 4.8 | >30 |

*Table 1. Performance comparison between our work and other ring-based LN resonators.*

*4. Conclusion*

In summary, the thin film LN ring-pair modulator has demonstrated superior modulation performance over the single-ring modulator. Without increasing the optical loss, the ring-pair modulator has 1.78-times larger linewidth (22 GHz) than that of single-ring modulator under the same ring parameters. The extinction ratio also adds up to 32 dB in ring-pair modulator, which paves a new approach to high extinction ratio ring-based modulators without critical coupling requirements. Furthermore, the ring-pair modulator inherits ultra-compact size of ring modulators, which is attractive for large-scale functional photonic integration. Owing to its larger working wavelength window, higher extinction ratio with low optical loss and compact size, our high-performance thin film LN ring-pair modulator provides new building path to open up the photonic applications of ring modulators.

*5. Methods*

Device fabrication: Devices with 1.5 um waveguides are fabricated from a commercially available x-cut thin film LN (nanoLN), in which a 600 nm thin film LN sits on top of a 2 um silicon dioxide on a silicon wafer. Ring resonators and bus waveguide are first defined in a negative photoresist (FOX 16) by electron beam lithography (EBL, Elionix, 100 kV) and then transferred into LN thin film using Ar+ plasma-coupled reactive ion etching, similar with our previous work[38, 39]. 400 nm LN is etching, leaving a 200 nm slab. PMMA/MMA double layer photoresists are used with EBL overlayer alignment to define the electrodes patterns. The electrode gap is designed as 4.5 um. High precision alignment between photoresists and waveguide layers are required. The developed double layer structure is then used in lift-off process to generate 300 nm thick gold electrodes. The waveguides and electrodes are cladded with 1.5 um silicon dioxide layer using plasma enhanced chemical vapor deposition (PECVD).

A new photoresist S1813 is used with maskless aligner (MLA 150), and then treated with buffered oxide etch (BOE) wet etching to open windows for via on cladding layer. Another lithography and lift-off process are performed to produce the top connecting metal strips and probe pads. Thanks to the isotropic etching of BOE, the long gentle slope on the sidewall of open windows guarantees good electrical connecting between bottom and top metal pads. The chip is finally cleaved to make sure good optical coupling between fibers and waveguides.

Device characterization: The fabricated devices are evaluated with a home built optical analysis system which consists of a tunable laser, a polarization controller, and an optical spectrum analyzer. The tunable laser source (Santec TSL 510) works with the wavelength range from 1500 nm to 1630 nm. The transverse-electric mode excitation is ensured by a polarization controller and the light is coupled on/off chip through tapered single mode lens fibers and finally received by detectors.

**Conflict of Interest**

The authors declare no conflict of interest.

**References**

[1] G. T. Reed, G. Mashanovich, F. Y. Gardes, D. J. Thomson, *Nature Photonics* **2010**, 4, 518.
[2] D. A. B. Miller, *Journal of Lightwave Technology* **2017**, 35, 346.
[3] D. Zhu, L. Shao, M. Yu, R. Cheng, B. Desiatov, C. Xin, Y. Hu, J. Holzgrafe, S. Ghosh, A. Shams-Ansari, *Advances in Optics and Photonics* **2021**, 13, 242.
[4] T. M. Fortier, M. S. Kirchner, F. Quinlan, J. Taylor, J. C. Bergquist, T. Rosenband, N. Lemke, A. Ludlow, Y. Jiang, C. W. Oates, S. A. Diddams, *Nature Photonics* **2011**, 5, 425.
[5] Y. Xu, A. A. Sayem, L. Fan, C.-L. Zou, S. Wang, R. Cheng, W. Fu, L. Yang, M. Xu, H. X. Tang, *Nature Communications* **2021**, 12, 4453.
[6] J. Lu, M. Li, C.-L. Zou, A. Al Sayem, H. X. Tang, *Optica* **2020**, 7, 1654.
[7] J. Shang, H. Chen, Z. Sui, Q. Lin, K. Luo, L. Yu, W. Qiu, H. Guan, Z. Chen, H. Lu, *Optics Express* **2022**, 30, 14530.
[8] Y. Wang, S. Zhou, D. He, Y. Hu, H. Chen, W. Liang, J. Yu, H. Guan, Y. Luo, J. Zhang, *Opt. Lett.* **2016**, 41, 4739.

[9] S. Y. Siew, B. Li, F. Gao, H. Y. Zheng, W. Zhang, P. Guo, S. W. Xie, A. Song, B. Dong, L. W. Luo, C. Li, X. Luo, G. Q. Lo, *Journal of Lightwave Technology* **2021**, 39, 4374.

[10] M. Smit, K. Williams, J. v. d. Tol, *APL Photonics* **2019**, 4, 050901.

[11] E. L. Wooten, K. M. Kissa, A. Yi-Yan, E. J. Murphy, D. A. Lafaw, P. F. Hallemeier, D. Maack, D. V. Attanasio, D. J. Fritz, G. J. McBrien, *IEEE Journal of selected topics in Quantum Electronics* **2000**, 6, 69.

[12] A. Honardoost, K. Abdelsalam, S. Fathpour, *Laser & Photonics Reviews* **2020**, 14, 2000088.

[13] C. Haffner, D. Chelladurai, Y. Fedoryshyn, A. Josten, B. Baeuerle, W. Heni, T. Watanabe, T. Cui, B. Cheng, S. Saha, *Nature* **2018**, 556, 483.

[14] Y. Jia, J. Wu, X. Sun, X. Yan, R. Xie, L. Wang, Y. Chen, F. Chen, *Laser & Photonics Reviews* **2022**, 2200059.

[15] A. Prencipe, M. A. Baghban, K. Gallo, *ACS Photonics* **2021**, 8, 2923.

[16] S. Saravi, T. Pertsch, F. Setzpfandt, *Advanced Optical Materials* **2021**, 9, 2100789.

[17] R. Zhuang, J. He, Y. Qi, Y. Li, *Adv Mater* **2022**, e2208113.

[18] Y. Kong, F. Bo, W. Wang, D. Zheng, H. Liu, G. Zhang, R. Rupp, J. Xu, *Advanced Materials* **2020**, 32, 1806452.

[19] C. Wang, M. Zhang, X. Chen, M. Bertrand, A. Shams-Ansari, S. Chandrasekhar, P. Winzer, M. Loncar, *Nature* **2018**, 562, 101.

[20] M. Zhang, B. Buscaino, C. Wang, A. Shams-Ansari, C. Reimer, R. Zhu, J. M. Kahn, M. Loncar, *Nature* **2019**, 568, 373.

[21] Y. Jia, L. Wang, F. Chen, *Applied Physics Reviews* **2021**, 8, 011307.

[22] X. Han, Y. Jiang, A. Frigg, H. Xiao, P. Zhang, T. G. Nguyen, A. Boes, J. Yang, G. Ren, Y. Su, *Laser & Photonics Reviews* **2022**, 16, 2100529.

[23] X. Ye, S. Liu, Y. Chen, Y. Zheng, X. Chen, *Opt. Lett.* **2020**, 45, 523.

[24] J. Lin, F. Bo, Y. Cheng, J. Xu, *Photonics Research* **2020**, 8, 1910.

[25] Z. Huang, K. Luo, Z. Feng, Z. Zhang, Y. Li, W. Qiu, H. Guan, Y. Xu, X. Li, H. Lu, *Science China Physics, Mechanics & Astronomy* **2022**, 65, 1.

[26] M. Li, J. Ling, Y. He, U. A. Javid, S. Xue, Q. Lin, *Nature Communications* **2020**, 11, 1.

[27] M. Xu, M. He, H. Zhang, J. Jian, Y. Pan, X. Liu, L. Chen, X. Meng, H. Chen, Z. Li, *Nature communications* **2020**, 11, 1.

[28] G. Poberaj, H. Hu, W. Sohler, P. Guenter, *Laser & photonics reviews* **2012**, 6, 488.

[29] A. Boes, B. Corcoran, L. Chang, J. Bowers, A. Mitchell, *Laser & Photonics Reviews* **2018**, 12, 1700256.

[30] M. He, M. Xu, Y. Ren, J. Jian, Z. Ruan, Y. Xu, S. Gao, S. Sun, X. Wen, L. Zhou, *Nature Photonics* **2019**, 13, 359.

[31] Q. Xu, B. Schmidt, J. Shakya, M. Lipson, *Optics express* **2006**, 14, 9431.

[32] J. Sharma, H. Li, Z. Xuan, R. Kumar, C.-M. Hsu, M. Sakib, P. Liao, H. Rong, J. Jaussi, G. Balamurugan, presented at *2021 Symposium on VLSI Circuits*, **2021**.

[33] R. A. Cohen, O. Amrani, S. Ruschin, *Nature Photonics* **2018**, 12, 706.

[34] I. Krasnokutska, J.-L. J. Tambasco, A. Peruzzo, *Scientific Reports* **2019**, 9.

[35] J. Lu, J. B. Surya, X. Liu, A. W. Bruch, Z. Gong, Y. Xu, H. X. Tang, *Optica* **2019**, 6, 1455.

[36] L. Chen, Q. Xu, M. G. Wood, R. M. Reano, *Optica* **2014**, 1, 112.

[37] A. Rao, A. Patil, J. Chiles, M. Malinowski, S. Novak, K. Richardson, P. Rabiei, S. Fathpour, *Optics Express* **2015**, 23, 22746.

[38] I. Briggs, S. Hou, C. Cui, L. Fan, *Opt Express* **2021**, 29, 26183.

[39]    P. K. Chen, I. Briggs, S. Hou, L. Fan, *Opt Lett* **2022**, 47, 1506.